\documentstyle[aps,epsfig]{revtex}
\begin{document}
\title{ Generalized mass formula for non-strange and hyper nuclei with SU(6) symmetry breaking }

\author{C. Samanta$^{1,2}$ \thanks{E-mail:chhanda.samanta@saha.ac.in; csamanta@vcu.edu}, P. Roy Chowdhury$^1$}
\address{ $^1$ Saha Institute of Nuclear Physics, 1/AF Bidhan Nagar, Kolkata 700 064, India }
\address{ $^2$ Physics Department, Virginia Commonwealth University, Richmond, VA 23284-2000, U.S.A. }

\author { D.N. Basu}
\address {Variable  Energy  Cyclotron  Centre,  1/AF Bidhan Nagar, Kolkata 700 064, India}
\date{\today }
\maketitle
\begin{abstract}

     A simultaneous description of non-strange nuclei and  hypernuclei is provided by a 
single mass formula inspired 
by the spin-flavour SU(6) symmetry breaking. This formula is used to estimate the hyperon 
binding energies of Lambda, double Lambda, Sigma, Cascade and Theta hypernuclei. 
The results are found to be in good agreement with the available
experimental data on 'bound' nuclei and relativistic as well as quark mean field calculations. This mass formula is useful 
to estimate binding energies over a wide range of masses including the light mass nuclei. It is not applicable for repulsive potential.


\end{abstract}
\vskip 0.2cm

      The hypernuclear physics is of great importance in many branches of physics. Of particular interest is the 
understanding of strange particles in baryonic matter, since many questions in heavy-ion physics, particle physics
and astrophysics are related to the effect of strangeness (S) in nuclear matter. Moreover, the contribution of the 
hyperons strongly influences the mass of neutron stars as well. In the past decade considerable amount of spectroscopic 
informations were accumulated experimentally on the $\Lambda^0$ (S=-1)hypernuclei. The $\Lambda$ separation energies were 
determined for the ground states of about 40 $\Lambda$ hypernuclei including several double-$\Lambda$ 
hypernuclei \cite{Ba90}, \cite{Ta01}. 
Doubly strange hypernuclei also arise in a form of $\Xi^-$(S=-2) hypernuclei and were studied both 
theoretically \cite{Do93} and experimentally \cite{Ba90}, \cite{Fu98}, \cite{Kh00} in a limited number of nuclei. 
Sigma hyperons exhibit interesting property in its interaction with nuclei. Several studies suggest that due to 
strongly repulsive $\Sigma$-nucleus potential Sigmas are unbound in nuclei, except for the very special case of 
nuclei with mass number A=4 \cite{Saha04}. Existance of a T= 1/2, S= 0 bound state in $^4_\Sigma He$ was first 
predicted by Harada et al. \cite{Hara90} and has been found 
experimentally \cite{Naga98}, \cite{Outa94}, \cite{Haya92}, \cite{Sa98}. The $1/A$ dependence of the Lane term 
 present  in the isospin-dependent $\Sigma$-nucleus potential \cite{Hara90} reduces the possibility of finding 
bound $\Sigma$-hypernuclei with large A.

   In this scenario, an exotic hyperon  $\Theta^+$ (S=+1, mass $\sim$1530 MeV,
width $<$15 MeV ) with exotic pentaquark structure was predicted in 1997 \cite{Di97}. 
Announcement of its discovery at Spring-8, Japan \cite{Na03} sparked an avalanche 
of activities in the field of hyperons and hypernuclei \cite{ZC05}. Now a question has arisen 
whether the $\Theta^+$  hyperon really exists \cite{Cl05} and if so, whether it will be bound 
in a nucleus \cite{Mi04}, \cite{Ca05}.  
Calculations in a relativistic mean-field formalism (RMF) suggest that 
as there is an attractive $\Theta^+$-nucleus interaction, the $\Theta^+$ particle can be 
bound in nuclei and, the $\Theta^+$  hypernuclei would be bound more strongly than $\Lambda$ 
hypernuclei \cite{Zh05}. Recent calculations in quark mean-field (QMF) model also support 
existence of bound $\Theta^+$ hypernuclei and predict that in comparison to $\Lambda$ 
hypernuclei more bound states are there in $\Theta^+$ hypernuclei \cite{Sh05}.
While a search for bound Theta hypernuclei is on, for a large number of 
hypernuclei, including double Lambdas, Cascade and Sigmas, more experimental data are needed. One also needs
to have an apriori estimation of their possible binding energies in a wide mass range for planning of the experiment 
and to locate the peaks in the experimental missing mass spectra.

      Although the $\Lambda$, $\Sigma$, $\Xi$ and $\Theta$ hyperons are all baryons, no single 
mass formula exists which can predict 
the binding energies of all of them on the same footing as the non-strange nuclei. In this work 
we present a generalized mass formula 
for both the strange and non-strange nuclei which does not disregard the normal nuclear matter 
properties. 
It is a straight forward equation and binding energy of non-strange as well as bound hypernuclei 
can be estimated by using this 
single equation. Its predictions compare well with the available experimental data. However it is not applicable 
for repulsive potential.

      Earlier, Dover et al \cite{Do93} prescribed two separate mass formulae for $\Lambda$ and 
$\Xi$ hypernuclei by introducing several volume and symmetry terms in Bethe-Weizs\"acker mass 
formula (BW).
The BW formulae given in Ref. \cite{Do93} were developed for a much broader context (multi-Lambda systems or strange 
hadronic matter) with parameters inspired on the strengths of the YN and YY interactions. In this respect, 
the single-Lambda BW equation given in \cite{Do93} is a limit of a more general formula. The BW equation proposed in 
Ref. \cite{Le98}  is inspired by the spin-flavour SU(6) symmetry and the pairing term is replaced by the expectation 
value of the space-exchange or, Majorana operator. The Majorana operator in the ground state configurations of 
single- and double-Lambda hypernuclei were expressed in terms of N, Z, and Lambda number. A strangeness dependent 
symmetry breaking term was also incorporated. This prescription gave reasonable description of the experimental data 
on single-and double-Lambda hypernuclei separation energies.
As none of these formulations had explicit hyperon mass consideration, they can not be used for binding 
energy calculation of other hypernuclei. Both the formulae have some other limitations which 
will be discussed later.

      The Wigner's $SU(4)$ symmetry arises as a result of the combined invariance in spin (I) 
and isospin (T). In order to incorporate the strangeness degree of isospin, $SU_T(2)$ is 
replaced by $SU_F(3)$ and the combined spin(I)-flavour(F) invariance gives rise to the  
$SU(6)$ classification of Gursey and Radicati \cite{Gu64}. The $SU_F(3)$ symmetry breaks by 
explicit consideration of a mass dependent term in a mass formula. The $SU(6)$ symmetry 
breaking is related to different strengths of the 
nucleon-nucleus and hyperon-nuclear interactions and has important
consequences. For example, although small, the $\Sigma-\Lambda$ mass 
difference figures prominently in the smallness of the $\Lambda$-nuclear spin-orbit 
interaction \cite{Ka05}. 

      In this work the non-strange normal nuclei and strange hypernuclei are treated on the same
footing with due consideration to SU(6) symmetry breaking. 
The generalization of the mass formula is pursued starting from the modified-Bethe-Weizs\"acker 
mass formula (BWM) preserving the normal nuclear matter properties. The BWM is basically the 
Bethe-Weizs\"acker mass 
formula extended for light nuclei \cite{Sa02}, \cite{Ad04}, \cite{Cs04}, \cite{SA04} which can 
explain the gross properties 
of binding energy versus nucleon number curves of all non- strange normal nuclei from Z=3 to 
Z=83. In BWM, the binding energy of a nucleus of mass number A and 
total charge Z is defined  as      

\begin{equation}
B(A,Z) = a_vA-a_sA^{2/3}-a_cZ(Z-1)/A^{1/3}-a_{sym}(N-Z)^2/[(1+e^{-A/k})A]+\delta_{new},
\label{seqn1}
\end{equation}
\noindent
where for normal nuclei $N$ and $Z$ are the number of neutrons and protons respectively and
\begin{equation}
a_v=15.777~MeV,~a_s=18.34~MeV,~a_c=0.71~MeV,~a_{sym}=23.21~MeV~and~k=17,
\label{seqn2}
\end{equation}
\noindent
and the pairing term,

\begin{equation}
\delta_{new}=(1-e^{-A/c})\delta,~~where~~c=30,
\label{seqn3}
\end{equation}
\noindent
and

\begin{eqnarray}
 \delta=&&12A^{-1/2}~for~even~neutron-even~proton~number,~=-12A^{-1/2}~for~ odd~neutron-odd~proton~number, \nonumber\\
 &&~=0~ when~ total~ neutron~ plus~ proton~ number~ is~ odd.\nonumber\\
\label{seqn4}
\end{eqnarray} 
\noindent

Hypernuclei are found to be more bound than normal nuclei. 
A systematic search of experimental data of hyperon separation energy ($S_Y$) for $\Lambda$, 
$\Lambda \Lambda$, $\Sigma^0$ and $\Xi^-$ hypernuclei  
leads to a generalised mass formula for hyper and non-strange nuclei which will be, henceforth, called the BWMH. 
The experimental  $S_Y$ of the $\Lambda$-hypernuclei (for which experimental data are available over a wide 
mass range) is found to follow a relation $S_Y \propto A^{-2/3}$ \cite{Ch88}. The available data  for Cascade 
hypernuclei also follow a similar $S_Y \propto A^{-2/3}$ trend but with different slope. 
Unlike Levai et al. \cite{Le98}, the SU(6) symmetry breaking term represented by the 
strangeness is taken here to be inversely proportional to $A^{2/3}$. 
Explicit inclusion of the pairing term partly accounts for the Majorana term while 
still preserving the nuclear 
saturation properties. An additional mass dependent term breaks the $SU_F(3)$ symmetry.

In BWMH the hypernucleus is considered as a core of normal nucleus plus 
the hyperon(s) and the binding energy is defined as 

\begin{eqnarray}
B(A,Z) = &&a_vA-a_sA^{2/3}-a_cZ(Z-1)/A^{1/3}-a_{sym}(N-Z_c)^2/[(1+e^{-A/k})A]+\delta_{new} \nonumber\\
              && + n_Y [c_0 . (m_Y) - c_1 - c_2 \mid S \mid / A^{2/3}], \nonumber\\
\label{seqn5}
\end{eqnarray}
\noindent
where $n_Y$ = number of hyperons in a nucleus, $m_Y$ = mass of the hyperon in 
$MeV$, $S$ = strangeness of the hyperon and mass number $A = N + Z_c + n_Y$ is equal to the 
total number of baryons. $N$ and $Z_c$ are the number of neutrons and protons respectively 
while the $Z$ in eqn.(5) is  given by     
\begin{equation}
Z = Z_c + n_Y q
\label{seqn6}
\end{equation}
\noindent
where $q$ is the charge number (with proper sign) of hyperon(s) constituting the hypernucleus.
For non-strange (S=0) normal nuclei, $Z_c = Z$ as $n_Y$ =0. 
The choice of $\delta_{new}$ value depends on the number of neutrons and protons in both normal
and hypernuclei.   
For example, in case of $^{9}_{\Lambda}Li$ the neutron number N=5(odd), proton number $Z_c$=3(odd),  
and $n_Y$=1. Therefore, $\delta=-12A^{-1/2}$ as the (N, $Z_c$) combination is odd-odd, 
although the total baryon number A= $A = N + Z_c + n_Y$ =9(odd). Whereas, for non-strange normal $^{9}Li$ nucleus 
$\delta=0$ for A=9(odd).

 In eqn.(5), the constants $c_0, c_1$ and $c_2$ have been fixed from an empirical fit to  the experimental values of $S_Y$ for thirty-five $\Lambda$, 
three $\Lambda-\Lambda$ and six $\Xi^-$ hypernuclei (Table 1). 

\begin{table}
\caption{Experimental data on hyperon separation energies ($S_Y$) of hypernuclei with experimental errors ($\Delta S_Y)$,
 number of hyperons ($n_Y$) and $NS$= $n_Y . S$ where S is the strangeness.}

\begin{tabular}{cccccc}
Hypernuclei &$S_Y$ &$\Delta S_Y$& $n_Y$ & $NS$ & Ref.     \\ 
&MeV&MeV&&\\ \hline
&&&&& \\
$^{4}_{\Lambda}H$& 2.04& 0.04& 1& -1& \cite{Ba90} \\
&&&& \\
$^{4}_{\Lambda}He$& 2.39& 0.03& 1& -1& \cite{Ba90} \\
&&&& \\
$^{5}_{\Lambda}He$& 3.12& 0.02& 1& -1& \cite{Ba90} \\
&&&& \\
$^{6}_{\Lambda}He$& 4.18& 0.10& 1& -1& \cite{Ba90} \\
&&&& \\
$^{7}_{\Lambda}He$& 5.23&  0.00& 1& -1& \cite{Maj95} \\
&&&& \\
$^{8}_{\Lambda}He$& 7.16& 0.70& 1& -1& \cite{Ba90} \\
&&&& \\
$^{6}_{\Lambda}Li$& 4.50& 0.00& 1&-1& \cite{Ba90} \\
&&&& \\
$^{7}_{\Lambda}Li$& 5.58& 0.03& 1& -1& \cite{Ba90} \\
&&&& \\
$^{8}_{\Lambda}Li$& 6.80& 0.03& 1& -1& \cite{Ba90} \\
&&&& \\
$^{9}_{\Lambda}Li$& 8.50& 0.12& 1& -1& \cite{Ba90} \\
&&&& \\
$^{7}_{\Lambda}Be$& 5.16& 0.08& 1& -1& \cite{Ba90} \\
&&&& \\
$^{8}_{\Lambda}Be$& 6.84& 0.05& 1& -1& \cite{Ba90} \\
&&&& \\
$^{9}_{\Lambda}Be$& 6.71& 0.04& 1& -1& \cite{Ba90} \\
&&&& \\
$^{10}_{\Lambda}Be$& 9.11& 0.22& 1& -1& \cite{Ba90} \\
&&&& \\
$^{9}_{\Lambda}B$& 8.29& 0.18& 1& -1& \cite{Ba90} \\
&&&& \\
$^{10}_{\Lambda}B$& 8.89& 0.12& 1& -1& \cite{Ba90} \\
&&&& \\
$^{11}_{\Lambda}B$&10.24&0.05& 1& -1& \cite{Ba90} \\
&&&& \\
$^{12}_{\Lambda}B$&11.37&0.06& 1& -1& \cite{Ba90} \\
&&&& \\
$^{12}_{\Lambda}C$&10.76&0.19& 1&-1& \cite{Ba90} \\
&&&& \\
$^{13}_{\Lambda}C$&11.69&0.12& 1& -1& \cite{Ba90} \\
&&&& \\
$^{14}_{\Lambda}C$&12.17&0.33& 1& -1& \cite{Ba90} \\
&&&& \\
$^{14}_{\Lambda}N$&12.17&0.00& 1& -1& \cite{Ba90} \\
&&&& \\
$^{15}_{\Lambda}N$&13.59&0.15& 1& -1& \cite{Ba90} \\
&&&& \\
$^{16}_{\Lambda}O$&12.50&0.35& 1& -1& \cite{Pile91} \\
&&&& \\
$^{17}_{\Lambda}O$&13.59,&0.0&  1& -1& \cite{Lala88} \\
&&&& \\
$^{28}_{\Lambda}Si$&16.00&0.29& 1& -1& \cite{Pile91} \\
&&&& \\
$^{32}_{\Lambda}S$&17.50&0.50& 1& -1& \cite{Ba90} \\
&&&& \\
$^{33}_{\Lambda}S$&17.96&0.00& 1& -1& \cite{Lala88} \\
&&&& \\
$^{40}_{\Lambda}Ca$&18.70&1.10& 1& -1& \cite{Pile91} \\
&&&& \\
$^{41}_{\Lambda}Ca$&19.24&0.00& 1& -1& \cite{Lala88} \\
&&&& \\
$^{51}_{\Lambda}V$&19.90&1.00& 1& -1& \cite{Lala94} \\
&&&& \\
$^{56}_{\Lambda}Fe$&21.00&1.50& 1& -1& \cite{Lala94} \\
&&&& \\
$^{89}_{\Lambda}Y$&22.10&1.60& 1& -1& \cite{Lala94} \\
&&&& \\
$^{139}_{\Lambda}La$&23.8&1.00& 1& -1& \cite{Hase96} \\
&&&& \\
$^{208}_{\Lambda}Pb$&26.5& 0.5& 1& -1& \cite{Hase96} \\
&&&& \\
$^{6}_{\Lambda \Lambda}He$&  7.25&0.19& 2& -2& \cite{Ta01} \\
&&&& \\
$^{10}_{\Lambda \Lambda}Be$&  17.7& 0.4& 2& -2& \cite{Ba90} \\
&&&& \\
$^{13}_{\Lambda \Lambda}B$& 27.5& 0.7& 2& -2& \cite{Aoki91,Dov91} \\
&&&& \\
$^{8}_{\Xi^-}He$&  5.90& 1.2& 1& -2& \cite{Ba90} \\
&&&& \\
$^{11}_{\Xi^-}B$& 9.2&  2.2& 1& -2& \cite{Ba90} \\
&&&& \\
$^{13}_{\Xi^-}C$& 18.1& 3.2& 1& -2& \cite{Ba90} \\
&&&& \\
$^{15}_{\Xi^-}C$& 16.0& 4.7& 1& -2& \cite{Ba90} \\
&&&& \\
$^{17}_{\Xi^-}O$& 16.0& 5.5& 1& -2& \cite{Ba90} \\
&&&& \\
$^{28}_{\Xi^-}Al$& 23.2& 6.8& 1& -2& \cite{Ba90} \\
&&&& \\
\end{tabular} 

The convention used in Table 1 (as well as in the text) is $^A_{Y^q}Z$ where Z is the net charge, Y is the hyperon type with charge q  and, A is the total number of baryons \cite{Ba90,Hara90}. There exists another convention in the literature in which one uses $^A_{Y^q}Z_c$ for denoting the hypernuclei, where  $Z_c$ is the number of protons. Therefore, according to this other convention, the hypernucleus $^4_{\Sigma^+} He$ of the present manuscript will be $^4_{\Sigma^+} H$. Also, $^{10}_{\Theta^+} Li $ would be $^{10}_{\Theta^+ }He$, and $^{11}_{\Theta^+} Be$ would be $^{11}_{\Theta^+} Li$. 
\end{table}

Here we calculate the root mean square deviation i.e.,  r.m.s. $(\sigma)$ for the hyperon separation energies where, $\sigma^2 ~=~ (1/N)\Sigma~[(S_Y)_{Th.}-(S_Y)_{Ex.}]^2$. Due to small number of data, simultaneous 3 parameter search does 
not yield any meaningful result. After many trial searches we fixed the value of $c_0=0.0335$ and made two parameter 
search using 35 $\Lambda$ hypernuclei data which yielded $c_1$=27.59 (45) and $c_2$= 48.6 (21) with r.m.s.=1.416 MeV. 
For better results from actual plotting of the graphs, $c_1$=26.7 was chosen and, keeping it fixed, a two parameter search 
with 35 $\Lambda$ hypernuclei data lead to $c_0$=0.0327 (4) and $c_2$=48.6 (21) with r.m.s. deviation 1.416 MeV. 
Finally we fixed $c_1$=26.7 and $c_2$=48.7  
and performed a one parameter search with 35 $\Lambda$ hypernuclei data. This leads to $c_0$=0.0327 (2) with 
r.m.s.=1.375 MeV. We fixed $c_0$=0.0335 which yielded best results for the two parameter fit. Finally, from the overall best
fit to all the data (r.m.s $\sim$ 1.4 MeV), we choose $c_0$=0.0335; $c_1$=26.7 and $c_2$=48.7 in eqn.(5).  

\begin{eqnarray}
B(A,Z) = &&a_vA-a_sA^{2/3}-a_cZ(Z-1)/A^{1/3}-a_{sym}(N-Z_c)^2/[(1+e^{-A/k})A]+\delta_{new} \nonumber\\
              && + n_Y [0.0335 m_Y - 26.7 - 48.7 \mid S \mid / A^{2/3}], \nonumber\\
\label{seqn7}
\end{eqnarray}
\noindent

The hyperon separation energy $S_Y$ defined as

\begin{equation}
S_Y = B(A,Z)_{hyper} - B(A-n_Y, Z_c)_{core}, 
\label{seqn8}
\end{equation}   
\noindent
is the difference between the binding energy of a hypernucleus and the binding energy of 
its non-strange core nucleus.

It is interesting to note
that this single equation yields the values of $S_Y$ in reasonable agreement with the available 
experimental data of all known bound hypernuclei. 
Fig.[1] shows plots of $S_Y$ versus A for $\Lambda$ and $\Lambda \Lambda$  hypernuclei 
which are in good agreement with the experimental data \cite{Ba90}, \cite{Ta01}. 
The recent discovery of the $^{10}_{\Lambda}Li$ nucleus points to a value of 
$S_Y \cong 10-12 MeV$ \cite{Saha05} where the 
BWMH predicts $S_{\Lambda} [^{10}_{\Lambda}Li] = 11.4 MeV$. The BWMH prediction of $S_Y$ for 
the $_{\Lambda\Lambda}^6He$ also agrees well [Fig.1(b)] with the recent experimental 
value $7.3 \pm 0.2 MeV$ of Takahashi et al. \cite{Ta01}.

Harada et al. investigated the structure of $^4_\Sigma He$ by the coupled-channel calculation between the $^3H + \Sigma^+$ and the 
$^3He + \Sigma^0$ channels and predicted the $\Sigma^+$ binding energy in $^4_\Sigma He$  to be $3.7$ to $4.6 MeV$ \cite{Hara90}. 
Available experimental binding energy values  are $4.4 \pm 0.3 \pm 1 MeV$ \cite{Naga98}, $2.8 \pm 0.7 MeV$ \cite{Outa94}, $4 \pm 1 MeV$ \cite{Haya92}. The BWMH predicts 
binding energy of $\Sigma^0$ ($m_Y = 1192.55 MeV$) and $\Sigma^+$ ($m_Y = 1189.37 MeV$) in $^4_{\Sigma^0}He$ 
(=$\Sigma^0$ + $^3_2 He$) as $2.69 MeV$ and $^4_{\Sigma^+}He$ (=$\Sigma^+$ + $^3_1 H$) as $1.6 MeV$ respectively.
Search for heavier $\Sigma$ hypernuclei has been carried out by several authors \cite{Saha04}, \cite{Naga98},
\cite{Outa94}, \cite{Haya92}, \cite{Sa98}, \cite{Bart99} without success.
P. K. Saha et al. studied inclusive ($\pi^-$, $K^+$) spectra on 
C, Si, Ni, In and Bi targets and concluded that a $\Sigma$-nucleus potential is strongly repulsive in such relatively heavy nuclei \cite{Saha04}. 
It has been suggested that the binding of the Sigma in the light hypernuclear species like $^4_\Sigma He$ is mainly due to a strong isovector term of the Sigma-nucleus potential which emphasizes the differences of the Sigma-N interaction in the isospin T=1/2 and the T=3/2 channels. This term is not present in the BWMH formula. Nevertheless, without changing any parameter, this single general equation (BWMH) reproduces the binding energy of the 'bound' $^4_\Sigma He$ hypernucleus. 
As mentioned before, the theoretical formulation suggests that
the binding energy of the light Sigma hypernuclear species is provided by the isospin dependent Lane term (associated to the isospin asymmetry of the core) not included in the BWMH formula. Thus the binding energy of BWMH obtained  for light hypernuclei comes from the isospin-averaged (i.e. isospin independent) term which provides attraction, while recent analysis on heavier Sigma hypernuclei (dominated by this isospin independent term of the Sigma-nucleus potential) suggest repulsion. For  $\Sigma^{0} + ^2H$, $\Sigma^{+} + ^2H$, and $\Sigma^{-} + ^2H$  hypernuclei the separation energies predicted by BWMH are -3.53 MeV, -4.62 MeV and -3.37 MeV respectively indicating that these light Sigma hypernuclei would be unbound,
even if the potential is attractive.
 In fact, so far no bound state of these hypernuclei could be found in the experiment. Indeed, further data on Sigma hypernuclei are necessary to determine more conclusively whether the Sigma feels attraction or repulsion.

Recent experimental data for $\Xi^-$ + $^{12} C$ and
$\Xi^-$ + $^{11} B$ are not conclusive and predict values in the range of 3 to 10 MeV depending on the $\Xi^-$ well 
depth \cite{Fu98} while, Khaustov et al.  suggests potential depth to be less than 20 MeV \cite{Kh00}.
In absence of any new conclusive data, we used the available data tabulated in Ref.~\cite{Ba90}. 
BWMH estimations for the $\Xi^-$ separation energies compare well with the 
experimental values (Fig. 2a). As no experimental data exists so far on the bound Theta hypernuclei,
the $\Theta^+$ separation energies are compared with the recent theoretical predictions  and the $S_Y$ of  
$\Theta^+$ are found to be close to the quark mean field (QMF) calculations \cite{Sh05}.

Analysis of each term of eqn.(8) reveals that the $\delta$ term has very little 
contribution (positive) in $S_Y$. The $a_{sym}$ term difference also contributes positive but, quite 
small  except at very high $\mid (N-Z_c) \mid$. The $a_s$ term difference (asd) is always negative and being more 
so at small A (Fig.3). For q=0, the Coulomb term difference (acd) contributes positive but very small for all A. 
For q=-1 it is positive and increases rapidly with $A$ while for q=+1, the acd  is negative and decreases 
rapidly with $A$. In the $\Theta^+$ plot [Fig. 2(b)] a maximum at low A and decrease in $S_Y$ 
values at large $A$ arise mainly due to increasing  contribution from the 
"strangeness term" (i.e., sterm1= $-48.7 \mid S \mid / A^{2/3}$), surface term (asd) and decreasing (negative) contribution
from the Coulomb term (fig.3c). For $q=0$ and $-1$ such maxima in Lambda plot 
[Fig. 1(a),1(b)] and $\Xi^-$ plot [Fig. 2(a)] are absent as asd, acd and sterm1 terms make increasing 
contributions with A (fig.3a, 3b).

Interestingly, the large masses of hyperons make some strongly bound 
hypernuclei whose non-strange core nuclei might be unbound. For example, BWMH shows that the prediction for 
the hyperon separation energy $S_Y$ for the $^{10}_{\Theta+}Li$ hypernucleus is 26.4
$MeV$ and that for the $^{11}_{\Theta+}Be$ hypernucleus is 25.1 $MeV$, the cores of which 
(i.e., $^{9}He$ and  $^{10}Li$ respectively) are known to be unbound. BWMH predicts
the $\Theta^{+} +^{12}C$
binding energy to be 23.24 MeV where as, the same for the 
$\Lambda +^{12}C$ binding energy is 12.65 MeV. This shows that compared to the 
$\Lambda$ hyperons, the  $\Theta^{+}$ hyperons will be more
strongly bound in a nucleus. This finding is in consonance with the prediction of 
Zhong et al. \cite{Zh05} and QMF calculation \cite{Sh05}.

      It is pertinent to note that unlike the previous two mass formulae \cite{Do93}, \cite{Le98} 
the BWMH mass formula does not jeopardize the normal nuclear matter properties. Fig.4 shows that the BWMH predicts the
line of stability quite well while the mass formula of Dover et al. \cite{Do93} shows mismatch 
for medium and heavy nuclei. The mass formula of Levai et al. \cite{Le98}
although reproduces the line of stability, it does not reproduce the binding energy 
versus mass number ($A$) plot (Fig. 5a) of non-strange light nuclei such as, $^6 Li$ \cite{Au03}. 
Mass formula of Dover et al. also shows marked deviations for isotopes with larger neutron 
numbers. The sharp oscillations in the experimental neutron separation 
energy ($S_n$) versus $A$  plots (Fig. 5b) are reproduced by BWMH but, not by the other two as
the pairing term is altogether absent in the mass formula of 
Dover et al. \cite{Do93} and, in case of the mass formula of Levai et al. \cite{Le98} it is too 
small. The incompressibility of infinite nuclear matter \cite{Ba04} obtained using the energy 
co-efficient $a_v$ of the volume term is about 300 MeV \cite{Bc04} for BWM (and BWMH) which is 
within the limits of 
experimental values while the mass formula of Levai et al. \cite{Le98} predicts values in the 
range of 400-480 MeV which 
are too high to be realistic. The presence of Majorana term in the mass formula of Levai et al. 
(eqns. (6) and (8) of \cite{Le98}) also poses a serious problem 
that the binding energy per nucleon diverges as mass number $A$ goes to infinity. 
This violates the nuclear saturation properties. As BWMH
is not plagued with such divergence problems and nuclear saturation properties are well 
preserved for large $A$, this mass formula will be useful for extension to astrophysics
related problems, like equation of state of a neutron star.

In summary, a single mass formula (BWMH) valid for both non-strange normal
nuclei and strange hypernuclei is prescribed by introducing hyperon mass
and strangeness dependent SU(6) symmetry breaking terms in the
modified-Bethe-Weizs\"acker mass formula (BWM). The BWMH preserves the
normal nuclear matter properties. Due to delicate balance amongst the
surface, strangeness and Coulomb contributions, a maximum in $q=+1$
hyperon separation energies is found near lower A values along with a
decreasing trend at larger A. This feature is absent in hypernuclei with
$q=0$ and $q=-1$ hyperon(s). Hyperon separation energies calculated by
BWMH are in good agreement with the available experimental data on 'bound'
nuclei, including $^4_\Sigma He$. Heavier Sigma-hypernuclei were predicted
to be unbound for repulsive Sigma-nucleus potential \cite{Saha04}. As BWMH
does not account for the repulsive potential, it predicts high binding
energies for the heavy Sigma hypernuclei which contradicts the present
wisdom. On the other hand, several authors have predicted that
Theta-hypernuclei would be bound more strongly than Lambda hypernuclei.
Quark mean field estimates for the Theta-separation energies are now
available for medium to heavy-mass Theta-hypernuclei \cite{Sh05} and those
predictions are in close agreement with the BWMH predictions. It is
noteworthy that BWMH can predict separation energies of both light and
heavy hyper-nuclei.
In view of the proposed search of the elusive $\Theta^+$ hypernuclei and
for many other hypernuclei which are predicted to be 'bound' (but their  binding energies  are unknown)
the present mass formula is expected to provide a useful guideline.
    
   We gratefully acknowledge G. Levi for sending us an useful compilation of some experimental data on hypernuclei.

\begin{figure}[h]
\eject\centerline{\epsfig{file=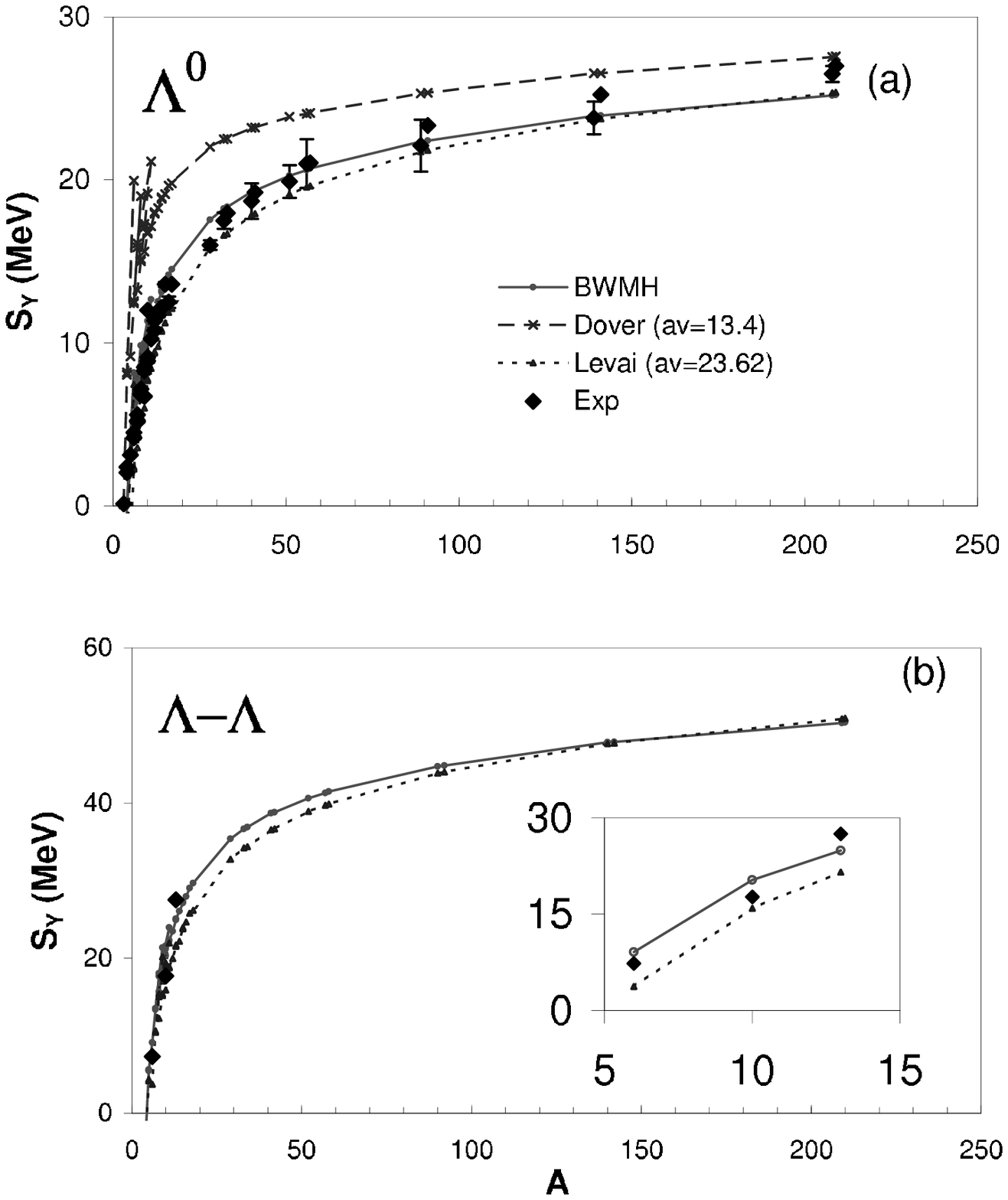,height=20cm,width=15cm, angle=0}}
\vskip 1.0cm
\caption
{ Hyperon separation energy $S_{Y}$ versus mass number $A$ for (a) single $\Lambda$
($m_Y$=1115.63 $MeV$) predicted by  BWMH (solid),
Dover et al. [3] (top most, dashed-line) and Levai et al. [20] (dotted) and experimental
values [1,~2] (rombus with error bars) and, 
(b) double $\Lambda$ predicted by  BWMH (solid),  and Levai et al. [20] (dotted) and experimental
values [1,~2] (rombus with error bars). In the inset the three
data points of double $\Lambda$ are shown with predictions of BWMH.
In all the figures, lines are added only as guides to the eyes. }
\label{fig1}
\end{figure}

\begin{figure}[h]
\eject\centerline{\epsfig{file=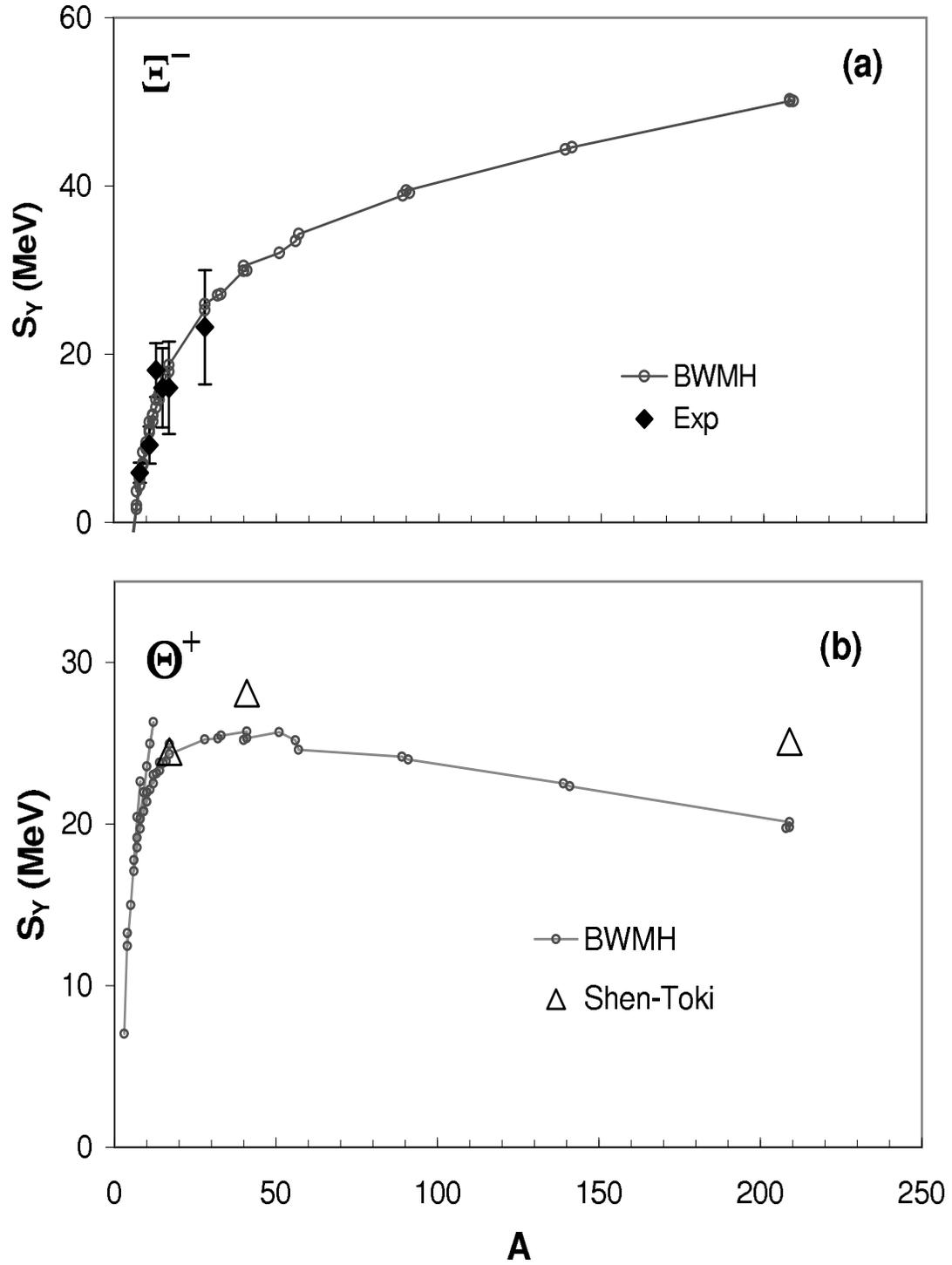,height=20cm,width=15cm, angle=0}}
\vskip 1.0cm
\caption
{ Hyperon separation energy $S_{Y}$ versus mass number $A$ for 
(a) single $\Xi^-$ ($m_Y$=1321.32 $MeV$) and experimental 
values [1] and, (b) single 
$\Theta^+$ ($m_Y$=1540 $MeV$) 
separation and quark mean field calculations of Shen and Toki [19].}
\label{fig2}
\end{figure}

\begin{figure}[h]
\eject\centerline{\epsfig{file=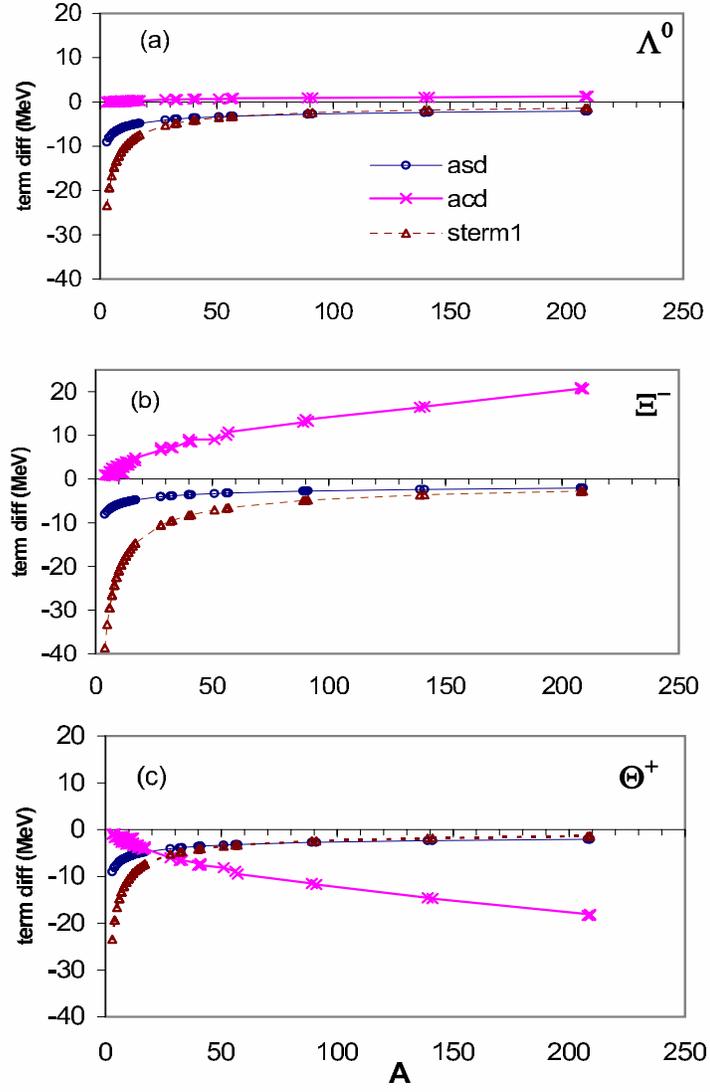,height=15cm,width=10cm, angle=0}}
\caption
{Contribution from Surface (asd), Coulomb (acd) and Strangeness (sterm1) term differences
to the hyperon separation energies (eqn.8) for
(a) $\Lambda^0$,
(b) $\Xi^-$ and (c) $\Theta^+$.}
\label{fig3}
\end{figure}

\begin{figure}[h]
\eject\centerline{\epsfig{file=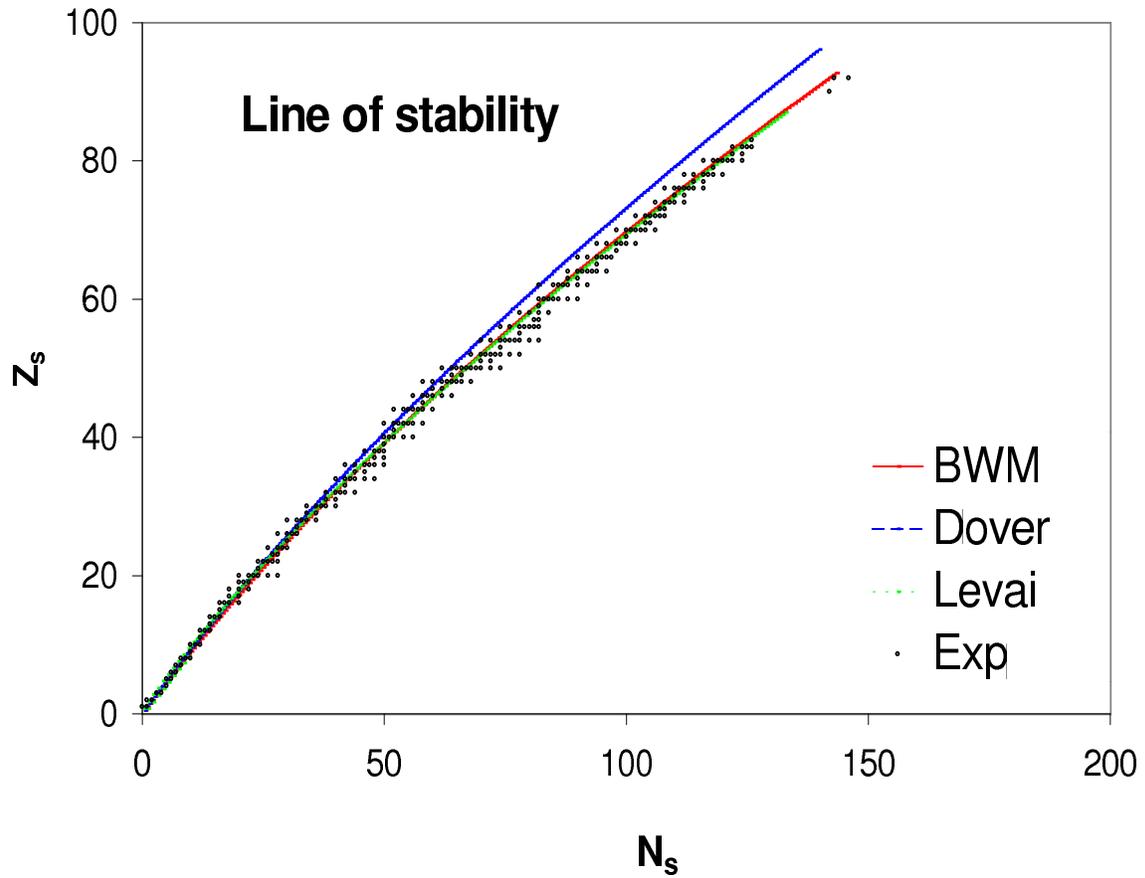,height=16cm,width=12cm, angle=-90}}
\vskip 1.0cm
\caption
{ The BWMH, Dover et al. [3] and Levai et al. [20] predictions for line of 
stability and experimentally observed stable nuclei [37]. Predictions of BWMH and Levai et al
are almost identical and those of Dover et al. deviate from the experimental values in heavier nuclei. }
\label{fig4}
\end{figure}

\begin{figure}[h]
\eject\centerline{\epsfig{file=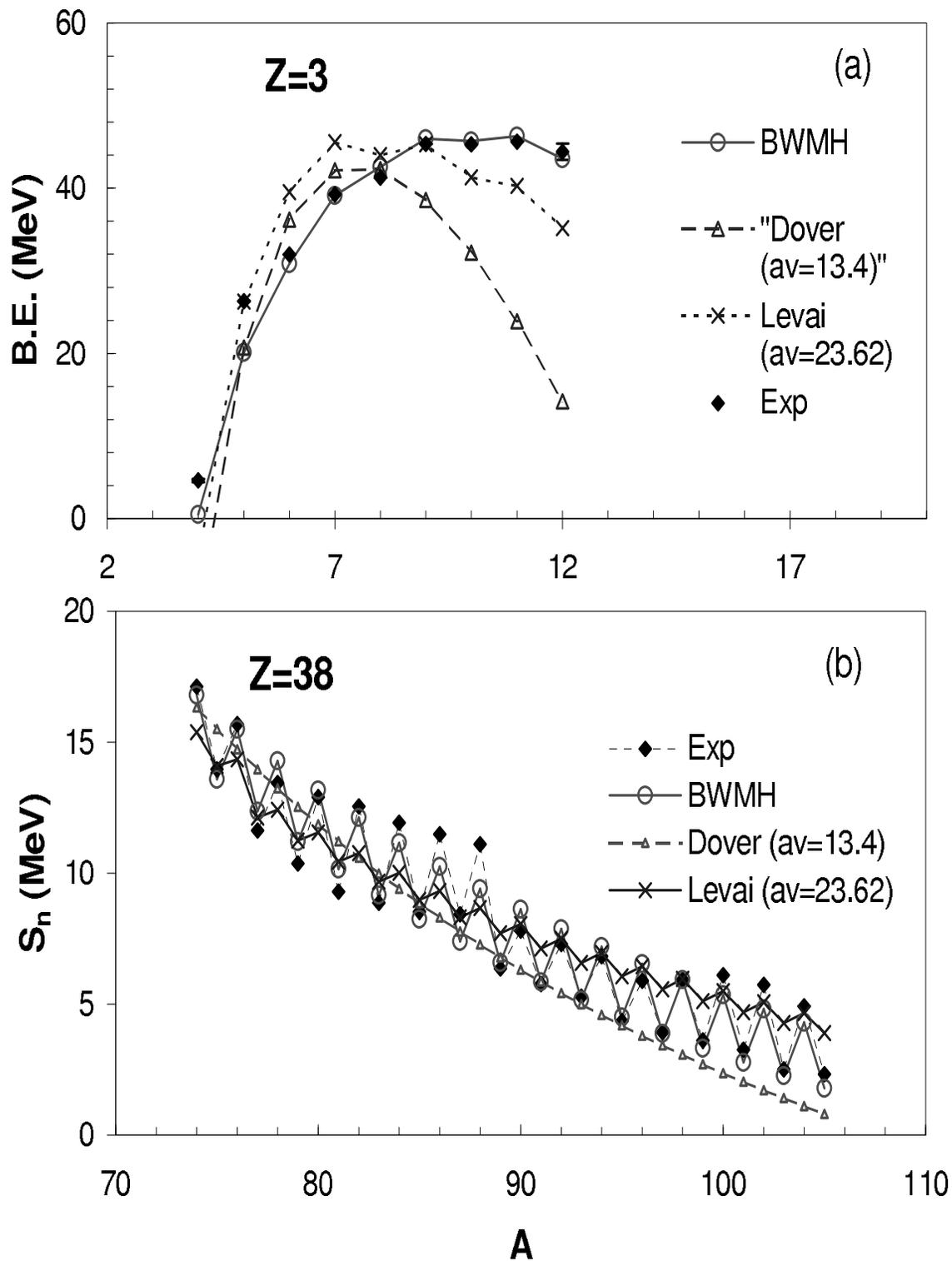,height=20cm,width=15cm, angle=0}}
\vskip 1.0cm
\caption
{ Predictions of BWMH, Dover et al. [3] and Levai et al. [20] 
and the experimental data [37] for (a) binding energy versus mass number $A$ for Z=3 and 
(b) the one neutron separation energy versus A for Z=38. }
\label{fig5}
\end{figure}

\end{document}